\documentclass[pra,twocolumn,showpacs,superscriptaddress]{revtex4}

\usepackage{graphicx}
\usepackage{amssymb}

\begin{document}

\title{Theoretical study of a cold atom beam splitter}

\author{Naceur Gaaloul}
\affiliation{Laboratoire de Photophysique Mol\'{e}culaire, CNRS, B\^{a}timent 210, Universit\'{e} Paris-Sud 11, 91405 Orsay cedex, France.}
\affiliation{Laboratoire de Spectroscopie Atomique, Mol\'eculaire et Applications, Department of Physics, Faculty of Sciences of Tunis, University Tunis El Manar, 2092 Tunis, Tunisia.}

\author{Annick Suzor-Weiner}
\affiliation{Laboratoire de Photophysique Mol\'{e}culaire, CNRS, B\^{a}timent 210, Universit\'{e} Paris-Sud 11, 91405 Orsay cedex, France.}

\author{Laurence Pruvost}
\affiliation{Laboratoire Aim\'{e} Cotton, CNRS, B\^{a}timent 505, Universit\'{e} Paris-Sud 11, 91405 Orsay cedex, France.}

\author{Mourad Telmini}
\affiliation{Laboratoire de Spectroscopie Atomique, Mol\'eculaire et Applications, Department of Physics, Faculty of Sciences of Tunis, University Tunis El Manar, 2092 Tunis, Tunisia.}

\author{Eric Charron}
\affiliation{Laboratoire de Photophysique Mol\'{e}culaire, CNRS, B\^{a}timent 210, Universit\'{e} Paris-Sud 11, 91405 Orsay cedex, France.}

\date{\today}

\begin{abstract}
A theoretical model is presented for the study of the dynamics of a cold atomic cloud falling in the gravity field in the presence of two crossing dipole guides. The cloud is splitted between the two branches of this laser guide, and we compare experimental measurements of the splitting efficiency with semi-classical simulations. We then explore the possibilities of optimization of this beam splitter. Our numerical study also gives access to detailed information, such as the atom temperature after the splitting.
\end{abstract}


\maketitle

\section{Introduction}
\label{sec:Intro}
The manipulation of cold atoms with optical fields is a very promising technique which is rapidly developping in the context of atom optics~\cite{Adams94}. Its applications range from laser cooling and trapping~\cite{Chu98,Cohen98,Phillips98,Wieman99} to coherent atom transport~\cite{Houde00,Kuhr01,Gustavson02,Dumke02,Wolschrijn02} and matter wave interferometry~\cite{Badurek88,Berman97}. Optical fields have also been proposed as an interesting tool to control the dynamics of internal and motional states of cold atoms for quantum information processing~\cite{Jaksch99,Hemmerich99,Brennen99,Calarco00,Charron02,Charron06}, and recent experimental studies have demonstrated a very promising first implementation with optical lattices~\cite{Mandel03,Porto03}.

For these types of applications, an effective way of guiding or transporting the atoms while keeping their coherence is required. For atom interferometry it is also necessary to separate the atomic wavefunction between the arms of an interferometer. Several experimental configurations have thus been explored for the implementation of an atom beam splitter with optical~\cite{Houde00,Dumke02} and magnetic~\cite{Muller00,Cassettari00,Muller01,Hommelhoff05} field potentials.

Recently, various configurations of atom beam splitters have been implemented and studied with coherent sources of atoms~\cite{Shin04,Wang05,Schumm05}. Indeed Bose-Einstein condensates are the best candidates for the implementation of cold atom interferometers because of their intrinsic coherence. However, for certain applications such as atomic clocks, gyrometers and gravimeters, thermal ensembles of cold atoms are still preferred because they allow for the preparation of a larger number of atoms at a higher repetition rate~\cite{Wieman99}. Furthermore, recent experiments~\cite{Miller05} show that high-contrast interference patterns with fringe visibility greater than 90\% can be obtained with thermal sources of  atoms.

In the case of dilute condensates, the dynamics is easily followed by solving the one-, two- or three-dimensional time-dependent Gross-Pitaevskii equation~\cite{Dalfovo99,Salasnich02}. With a thermal source of atoms, the theoretical description is more complex in the sense that many transverse modes of the guide are populated initially in an incoherent way. Implementing a cold atom interferometer with this type of atom source is therefore not trivial and requires a detailed theoretical investigation. In addition, new opportunities are now investigated with spatial light modulators~\cite{McGloin03} for the implementation of atom optics devices (guides, splitters,~\ldots) for thermal and coherent sources of atoms. The development of new theoretical models aimed at the description of the dynamics of a guided thermal ensemble of cold atoms is therefore needed for these applications.

A few theoretical studies of cold atom beam splitters have been published recently in various trapping situations~\cite{Kreutzmann04,Stickney03,Bortolotti04}. In these approaches it was assumed that the atomic wave packet was tightly confined in one dimension, and the effect of gravity was neglected. In the present article we study the cold atom beam splitter implemented in reference~\cite{Houde00}, by solving numerically the time dependent Schr\"odinger equation for the atomic motion in the presence of the gravity field and with realistic trapping potentials. Our aim is to propose a theoretical model which can reproduce the main features of this experiment, and to test its pertinence for more elaborate situations.

\begin{figure}[t!]
\centering
\includegraphics[width=8cm,angle=0,clip]{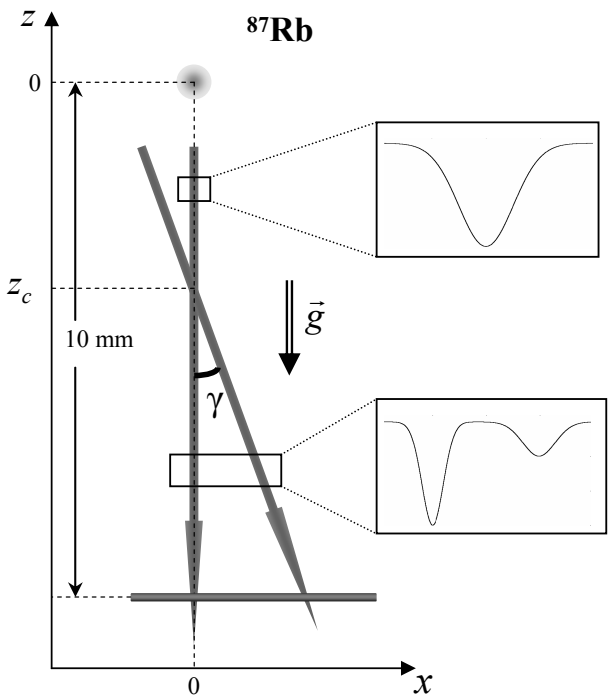}
\caption{Schematic view of the guiding setup. The initial $^{87}$Rb cloud is shown at $z=0$. When the magneto-optic trap is switched off, a vertical laser beam is switched on and the cloud, partially trapped by the associated dipole force, falls in the gravity field. At time $t_0\/$ a second oblique guide is switched on. The two guides cross at the height $z_c$ and form an \mbox{angle $\gamma$}. The trapping potential seen by the atoms at two different heights is shown in the insets. Finally, the atoms are probed 1~cm below the initial position.}
\label{fig:Schematic}
\end{figure}

A large ensemble of $^{87}$Rb atoms is initially trapped and cooled to a temperature $T_0$ in a magneto-optical trap (MOT) localized at the height \mbox{$z=0$} (see Figure~\ref{fig:Schematic} for a schematic view). At time \mbox{$t=0$}, the trapping potential is switched off, while a vertical far off-resonant laser beam, crossing the cloud close to its center, is switched on. The dipole interaction creates a potential well of depth $U_0$ which traps a significant portion of the atoms in the transverse directions $x$ and $y$.

The guided atoms then fall due to gravity, with a confined dynamics in the $x$ and $y$ directions. At time $t_0\/$, a second oblique guide is suddenly switched on. The two guides cross at the height $z_c$ and form an angle $\gamma$. A potential well, of depth $U_1$, is induced by the optical dipole interaction with the oblique guide. This creates an additional path for the motion of the atoms. Depending on the various parameters (light intensity, angle $\gamma$, temperature $T_0$, \ldots) a splitting of the cloud can be observed~\cite{Houde00} between the vertical and oblique branches.

\section{Theoretical Model}
\label{sec:Model}

In order to simplify the numerical treatment of this phenomenon, we restrict the dimensions of this study to the plane defined by the two guiding beams, {\em i.e.\/} to the $x$ and $z$ dimensions only. In addition, we adopt a semi-classical approach, where the effect of the gravity is treated classically. This approximation is justified by the value of the de Broglie wavelength associated with the speed of the particles in the $z$ direction, \mbox{$\lambda_{\rm db} \sim 1$~\AA}.

\subsection{Classical approach}

\begin{figure}[t!]
\centering
\includegraphics[width=8cm,angle=0,clip]{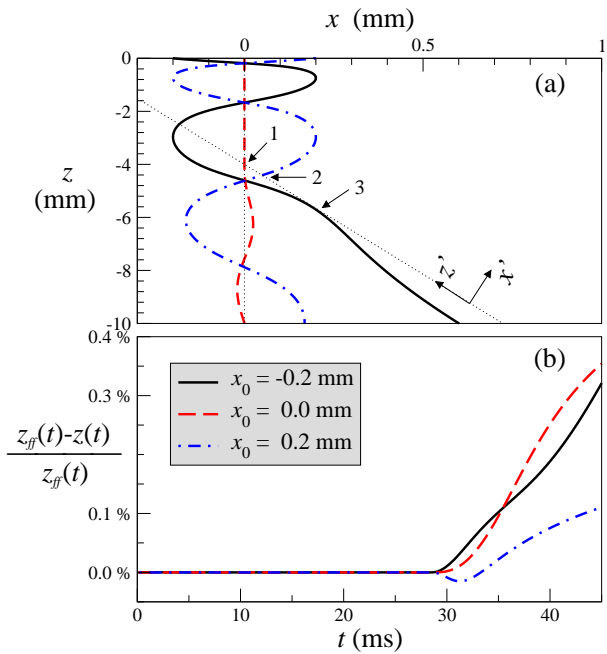}
\caption{(Color online) (a) Classical trajectories of a cold atom falling in the gravity field in the presence of the two trapping potentials with $U_0=30\,\mu$K, $U_1=10\,\mu$K, $w_0=0.2\,$mm and $w_1=0.3\,$mm. The three trajectories correspond to an initial height $z_0=0$ and initials positions $x_0=-0.2\,$mm (black solid line), $x_0=0$ (red dashed line) and $x_0=+0.2\,$mm (blue dash-dotted line) with zero initial momentum ($\dot{z}_0=0$ and $\dot{x}_0=0$). The two guides cross at the height $z_c=-4\,$mm with an angle $\gamma=0.12\,$rad, and the oblique guide is switched on at $t_0=28.6\,$ms. At this time, a free fall dynamics would give the position \mbox{$z_{f\!f}(t_0)=z_0-\dot{z}_0t-\frac{1}{2}gt_0^2=-4\,$mm} corresponding exactly to $z_c$. Unless specified, these laser parameters remain fixed throughout the paper. The thin vertical and oblique dotted lines reveal the geometry of the laser beams. (b) Deviation of the trajectories from a free fall dynamics in the $z$-direction for the same initial conditions as in (a).}
\label{fig:Classic}
\end{figure}

A two-dimensional classical trajectory \mbox{$\left\{x(t),z(t)\right\}$} is first evaluated by solving Newton's equations of motion for an atom initially at the position \mbox{$\left\{x_0,z_0\right\}$} with the momentum \mbox{$\left\{\dot{x}_0,\dot{z}_0\right\}$}. An efficient variable time-step Runge-Kutta integrator~\cite{rksuite} is used to solve these equations in the total potential \mbox{$V_{t}(x,z,t) = V_{g}(x,z,t) + mgz$}, where $m$ denotes the atomic mass and $g$ the gravitational constant. The guiding potential $V_{g}(x,z,t)$ is given by the following sum
\begin{equation}
\label{eq:Vguide}
V_{g}(x,z,t) = V_{0}(x) + V_{1}(x,z,t)\,,
\end{equation}
with
\begin{equation}
\label{eq:V01}
\left\{
\begin{array}{cclll}
V_{0}(x)     & = & -\,U_0 & {\rm e}^{-2\,x^2/w_{0}^{2} } \\
V_{1}(x,z,t) & = & -\,U_1 & u(t-t_0)\,\,{\rm e}^{-2\,x'^2/w_{1}^{2}}
\end{array}
\right.
\end{equation}
In the previous expressions $w_{0}$ and $w_{1}$ denote the waists of the vertical and oblique laser beams respectively, and $u(t-t_0)$ stands for the Heaviside step function. The following rotated coordinates
\begin{equation}
\label{eq:xpzp}
\left\{
\begin{array}{lcccccl}
x' & = & x       & \cos\gamma & + & (z-z_c) & \sin\gamma \\
z' & = & (z-z_c) & \cos\gamma & - & x       & \sin\gamma
\end{array}
\right.
\end{equation}
have also been introduced (see Figure~\ref{fig:Classic}a). The Gaussian form of this potential arises from the Gaussian intensity profile of the laser beam~\cite{Houde00}.

Figure~\ref{fig:Classic}a shows three typical trajectories with potential parameters close to the one chosen in the experiment performed in Orsay~\cite{Houde00}. From this graph it is already clear that, classically, only very specific initial conditions (such as $x_0=-0.2\,$mm and \mbox{$z_0=\dot{x}_0=\dot{z}_0=0$}) drive the atom in the oblique branch. Comparing the kinetic energy of the atom in the $x'$ direction at time $t=t_0$, $E_{x'}$, with the binding energy of the oblique guide gives the actual criterion which decides in favor or against the deviation of the atom from its natural vertical fall. This can be inferred from Figure~\ref{fig:Classic}a\,: the trajectory induced by the initial condition $x_0=+0.2\,$mm crosses the oblique guide almost perpendicularly at the point labelled 2 on the graph, with a maximum kinetic energy $E_{x'}$. This trajectory therefore remains almost unaffected by the presence of the oblique laser beam. On the contrary, the trajectory associated with the initial condition $x_0=-0.2\,$mm meets the oblique guide almost tangentially at the point labelled 3, and the atom is deflected from its initial vertical motion. The intermediate case $x_0=0$ follows a dynamics similar to the initial condition $x_0=+0.2\,$mm with a very slight deviation from the initial vertical motion. It will be shown in section~\ref{sec:Results} that this simple interpretation, in terms of individual trajectories, is not valid anymore when the atomic dynamics is treated at the quantum level.

Finally, Figure~\ref{fig:Classic}b shows the deviations of these trajectories from a simple free fall dynamics in the $z$-direction defined by the classical expression \mbox{$z_{f\!f}(t)=z_0-\dot{z}_0t-\frac{1}{2}gt^2$}. On the time interval \mbox{$0 \leqslant t \leqslant 45\,$ms}, required to reach the detection probe located at  $z_p=-10\,$mm, this deviation remains smaller than half a percent, and this justifies the approach adopted in the following, where the dynamics along the $z$-dimension is simply treated as a classical free fall. We now turn to the description of this semi-classical approach.

\subsection{Semi-classical treatment}

In the limit of dilute gases, the dynamics of the cloud can be simulated by solving the time-dependent Schr\"odinger equation along the $x$ and $z$ dimensions for the wavepacket $\Psi(x,z,t)$ describing the external dynamics of a trapped atom
\begin{equation}
\label{eq:TDSE2D}
i\hbar \frac{\partial}{\partial t} \Psi(x,z,t) =
\hat{\cal H}_{\rm 2D}(x,z,t)\; \Psi(x,z,t)\,.
\end{equation}
Since the initial state of the atomic wave packet can be described in general by a thermal mixture~\cite{Kreutzmann04}, the calculation of an observable at time $t$ can be done by a simple thermal average once $\Psi(x,z,t)$ is known (this averaging procedure is explained in section~\ref{sec:Results}, Eq.~(\ref{PRv0}-\ref{Psv0})). The two-dimensional Hamiltonian $\hat{\cal H}_{\rm 2D}(x,z,t)$ can be written as the following sum
\begin{equation}
\label{eq:Hamiltonian2D}
\hat{\cal H}_{\rm 2D}(x,z,t) = \hat{T}_x + \hat{T}_z + V_{g}(x,z,t) + m g z\,,
\end{equation}
where
\begin{equation}
\label{eq:Tq}
\hat{T}_q = - \frac{\hbar^2}{2m} \frac{\partial^2}{\partial q^2}
\end{equation}
denotes the kinetic energy operator along the $q$-coordinate.

The semi-classical approximation discussed above is introduced by replacing the $z$ coordinate in the two-dimensional Hamiltonian $\hat{\cal H}_{\rm 2D}(x,z,t)$ with the simple classical parameter \mbox{$z_{f\!f}(t)=z_0-\dot{z}_0t-\frac{1}{2}gt^2$}. The quantum dynamics of the atomic cloud along $x$ is then described by the time-dependent Schr\"odinger equation for the wavepacket $\varphi(x,t)$
\begin{equation}
\label{eq:TDSE}
i\hbar \frac{\partial}{\partial t} \varphi(x,t) =
\hat{\cal H}_{\rm 1D}(x,t)\; \varphi(x,t)\,,
\end{equation}
with a one-dimensional time-dependent Hamiltonian $\hat{\cal H}_{\rm 1D}(x,t)$ given by
\begin{equation}
\label{eq:Hamiltonian}
\hat{\cal H}_{\rm 1D}(x,t) = \hat{T}_x + V_{\rm 1D}(x,t)\,,
\end{equation}
where $V_{\rm 1D}(x,t)=V_{g}(x,z_{f\!f}(t),t)$. By adopting this one-dimensional approach, the numerical simulation is simplified at the cost of replacing the two-dimensional potential $V_{g}(x,z,t)$ (equations~(\ref{eq:Vguide}) and~(\ref{eq:V01})), which is time-independent for $t > t_0$, with the time-varying one-dimensional potential $V_{\rm 1D}(x,t)$. This potential changes slowly during the fall of the atom, and two of its snapshots are shown in the insets of Figure~\ref{fig:Schematic}.

\subsection{Time-dependent propagation}

We assume the atom to be initially in a well defined vibrational level $v_0$ of the vertical guide potential $V_0(x)$
\begin{equation}
\label{eq:initial_varphi}
\varphi(x,t=0)=\chi_{v_0}(x)\,,
\end{equation}
and we propagate the translational wavepacket until the time $t_f$ corresponding to the height of the detection probe, using the splitting operator method developed by Feit \emph{et al}~\cite{Feit83}
\begin{equation}
\label{eq:propage_varphi}
\varphi(x,t+\delta t) = {\rm e}^{-i\hat{\cal H}_{\rm 1D}\delta t/\hbar}\,\varphi(x,t)\,.
\end{equation}
The total Hamiltonian $\hat{\cal H}_{\rm 1D}(x,t)$ is splitted in two parts corresponding to the kinetic and potential propagators
\begin{equation}
\label{eq:split}
{\rm e}^{-i\hat{\cal H}_{\rm 1D}\delta t/\hbar} =
{\rm e}^{-i\hat{T}_x\delta t/2\hbar} \times
{\rm e}^{-iV_{\rm 1D}(x,t)\delta t/\hbar} \times
{\rm e}^{-i\hat{T}_x\delta t/2\hbar}
\end{equation}
to decrease the error to the order $(\delta t)^3$. The kinetic propagation is performed in the momentum space, and the potential propagation in the coordinate space. Fast Fourier Transformation (FFT) allows rapid passage back and forth from one representation to the other at each time step. Typical grids extend from $x_{min}=-1.0\,$mm to $x_{max}=2.0\,$mm with $N=2^{20}$ grid points, and a time step of the order of $\delta t \simeq 40\,\mu$s is used.

At the end of the propagation the wavefunction $\varphi(x,t_f)$ is analyzed to determine the efficiency of the beam splitter and to extract detailed information on the state of the atom in each branch of the laser guide.

\subsection{Initial trapping of the atomic cloud}

We assume the initial atomic cloud to be in a thermal state at temperature $T_0$ described by the usual Maxwell-Boltzmann phase-space probability distribution $W(x,z,\dot{x},\dot{z})$ defined by the following four-dimensional product
\begin{equation}
\label{WQQPP}
W(x,z,\dot{x},\dot{z}) =  W_Q(x) \times W_Q(z) \times W_{P}(\dot{x}) \times W_{P}(\dot{z})\,,
\end{equation}
where $W_Q(q)$ and $W_{P}(\dot{q})$ ($q=x,z$) are the position and momentum distributions
\begin{equation}
\label{eq:WQP}
\left\{
\begin{array}{cccl}
W_Q(q)         & = & \displaystyle\frac{1}{\sqrt{2\pi}\,\sigma_0}     &
{\rm e}^{-\,q^2/\,2\sigma_0^2 } \\[0.4cm]
W_{P}(\dot{q}) & = & \displaystyle\sqrt{\frac{m}{2\pi\,k_{\rm B}T_0}} &
{\rm e}^{-\,m\dot{q}^2/\,2k_{\rm B}\!T_0}
\end{array}
\right.
\end{equation}
In this expression $\sigma_0$ characterizes the size of the cloud and $k_{\rm B}$ is the Boltzmann constant. When the vertical guide is suddenly switched on at time $t=0$, only a fraction of the cold atoms are trapped in the dipole potential created by the laser beam intensity profile.

The total trapping probability $P_{\rm trap}$ can be calculated from the position and momentum distributions, assuming that for a given position $x$ the atom is trapped if its kinetic energy $E_x$ along this direction is lower than the binding energy $-V_0(x)$ of the potential. The integral over $x$ leads to the following rapidly convergent expansion
\begin{equation}
\label{Ptrap2}
P_{\rm trap} = \sum_{n=0}^{\infty}\left(-\frac{U_0}{k_{\rm B}T_0}\right)^n\,\frac{\beta_n}{(2n+1)\,n!}\,\,,
\end{equation}
where
\begin{equation}
\label{beta_n}
\beta_n = \frac{2\,w_0}{\displaystyle\left[w_0^2+(4n+2)\sigma_0^2\right]^{1/2}}\,\sqrt{\frac{U_0}{\pi\,k_{\rm B}T_0}}\,\,.
\end{equation}
This expression is the one-dimensional analogue of the well-known two dimensional probability given for example in~\cite{Wolschrijn02,Pruvost99,Houde_PhD02}. It should be noticed here that the trapping probability only depends on the two following dimensionless ratios $\sigma_0/w_0$ and $U_0/k_{\rm B}T_0$. This is a signature of the fact that the trapping probability can be expressed in a phase-space diagram as the overlap between the atomic cloud distribution and the trapping condition~\cite{Pruvost99,Houde_PhD02}.

\begin{figure}[t!]
\centering
\includegraphics[width=8cm,angle=0,clip]{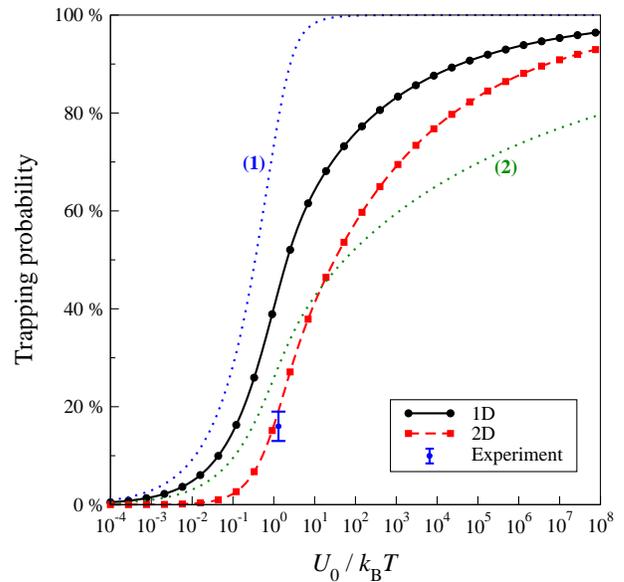}
\caption{(Color online) One-dimensional (black solid line) and two-dimensional (red dashed line) trapping probability as a function of the dimensionless ratio  $U_0/k_{\rm B}T_0$ (logarithmic scale) for $\sigma_0/w_0=1.5$. The blue point located at the abscissa $U_0/k_{\rm B}T_0=1.3$ with the error bar is an experimental measurement extracted from reference~\cite{Houde_PhD02}. The one-dimensional trapping probabilities for $\sigma_0/w_0=0.5$  and 2.5 are also shown as blue and green dotted lines with the labels (1) and (2) respectively. All other parameters are as in Figure~\ref{fig:Classic}.}
\label{fig:TrapProb}
\end{figure}

The variation of the trapping probability with the ratio $U_0/k_{\rm B}T_0$ is depicted in Figure~\ref{fig:TrapProb}. For a given $U_0$, the trapping probability changes very slowly with the temperature of the atomic cloud. In a real experiment, trapping occurs along both $x$ and $y$, and $P_{\rm trap}^2$ is therefore shown in Figure~\ref{fig:TrapProb} to compare with the measurement performed in~\cite{Houde_PhD02} for the ratio $U_0/k_{\rm B}T_0=1.3$. Finally, as one might intuitively guess, the trapping probability increases significantly when the size of the atomic cloud $\sigma_0$ decreases compared to the laser waist $w_0$ characterizing the size of the trapping potential.

Using arguments based on energy conservation, the probability for an atom to be trapped in a well defined initial vibrational state $v_0$ of total energy $\varepsilon_0$ can also be calculated using
\begin{equation}
\label{Pv0}
P(v_0) = \frac{1}{\rho(\varepsilon_0)}
\int_{-l_0}^{+l_0} W_Q(x)\,W_E\left(\varepsilon_0-V_0(x)\right)dx\,,
\end{equation}
once the density of states $\rho(\varepsilon_0)$ in the potential $V_0(x)$ is known. The positions $x=\pm\,l_0$, corresponding to \mbox{$2\,l_0^2=w_0^2\ln(-U_0/\varepsilon_0)$}, are the left and right turning points of the level $v_0$, and
\begin{equation}
\label{WE}
W_E(E)=\frac{\displaystyle{\rm e}^{-\,E\,/\,k_{\rm B}T_0}}{\displaystyle\sqrt{\pi\,E\,k_{\rm B}T_0}}
\end{equation}
is simply the analogue of $W_{P}(\dot{q})$, written in terms of kinetic energy \mbox{$E=\varepsilon_0-V_0(x)$}. The state-dependent probability $P(v_0)$ only depends on the three following dimensionless parameters: $\sigma_0/w_0$, $U_0/k_{\rm B}T_0$ and $\varepsilon_0/U_0$. It finally satisfies the relation
\begin{equation}
\label{SumPv0}
P_{\rm trap} = \sum_{v_0} P(v_0)\,.
\end{equation}

\begin{figure}[t!]
\centering
\includegraphics[width=8cm,angle=0,clip]{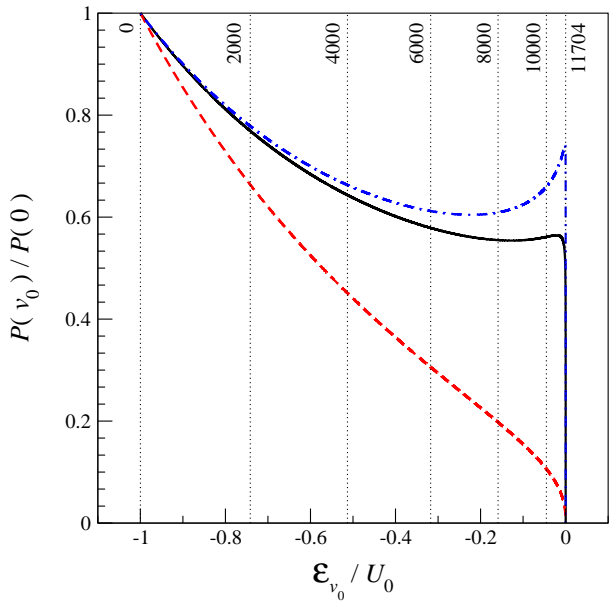}
\caption{(Color online) Initial population of the various vibrational levels $P(v_0)$ normalized with respect to the population of the ground state $P(0)$ as a function of the ratio of their energy $\varepsilon_0$ to $U_0$. The black solid, red dashed and blue dash-dotted lines correspond to $\sigma_0/w_0=1.5$, 0.5  and 2.5 respectively. All other parameters are as in Figure~\ref{fig:Classic}. The energies of the levels $v_0=0$, 2000, 4000, 6000, 8000, 10000 and 11704 are indicated by the thin vertical dotted lines.}
\label{fig:Prob_v}
\end{figure}

Figure~\ref{fig:Prob_v} shows the initial distribution of vibrational levels for various ratios $\sigma_0/w_0$. The lowest levels dominate the distribution, and this is particularly true when the size of the atomic cloud $\sigma_0$ is smaller than the size of the trapping potential $w_0$. Just as with the total trapping probability $P_{\rm trap}$ shown in Figure~\ref{fig:TrapProb}, $P(v_0)$ changes very slowly with the temperature. This variation is therefore not shown here.

\section{Numerical Results}
\label{sec:Results}
\begin{figure}[t!]
\centering
\includegraphics[width=8cm,angle=0,clip]{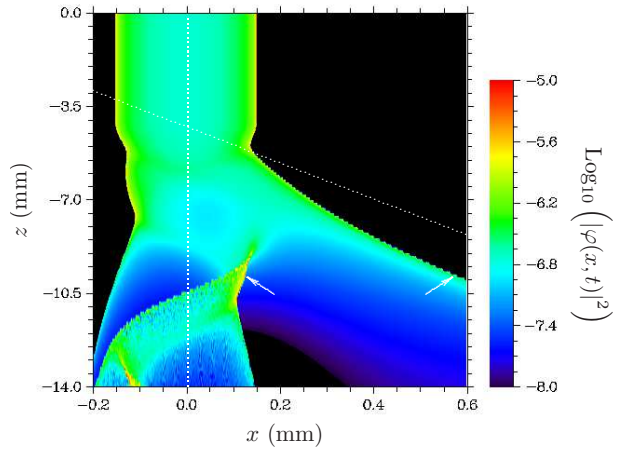}
\caption{(Color online) Contour plot depicting the time evolution of the envelope of $\left|\varphi(x,t)\right|^2$ on a logarithmic scale as a function of $x$ and of $z(t)=z_0-\dot{z}_0t-\frac{1}{2}gt^2$. The atom is falling in the gravity field in the presence of the two trapping potentials with $U_0=30\,\mu$K, $U_1=10\,\mu$K, $w_0=0.2\,$mm and $w_1=0.3\,$mm. The two guides cross at the height $z_c=-4\,$mm with an angle $\gamma=0.12\,$rad, and the oblique guide is switched on at $t_0=28.6\,$ms. The vertical and oblique dotted white lines reveal the directions of propagation of the laser beams. The two white arrows point to the location of the maximum probability density at the height $z=-10\,$mm. The initial level is here $v_0=6000$.}
\label{fig:Quantic}
\end{figure}

A typical quantum dynamics can be seen Figure~\ref{fig:Quantic}, which shows the time evolution of the initial level \mbox{$v_0=6000$} as a function of $x$ and $z$ for the initial condition $z_0=\dot{z}_0=0$. This initial state is a stationnary state of the vertical guide, and it does not evolve in time until it reaches the height $z_c=-4\,$mm where the two dipole guides cross. Afterwards, a wavepacket is created, and this one evolves inside two main branches, indicated by the white arrows. The oblique ``trajectory'' is guided by the oblique laser beam represented by the thin white oblique dotted line, while in the vertical branch an oscillating wavepacket is evolving. The picture obtained here is not as simple as the classical trajectories shown in Figure~\ref{fig:Classic} since this quantum state cannot be represented by a single trajectory. The consequence is that, depending on the initial conditions, the atomic wavepacket can be delocalized in the two guides simultaneously. A single initial quantum state can therefore split coherently along two paths separated by macroscopic distances. This effect, which is quantum by nature, might open some interesting perspectives for atom interferometry experiments with laser guides.

\begin{figure}[t!]
\centering
\includegraphics[width=8cm,angle=0,clip]{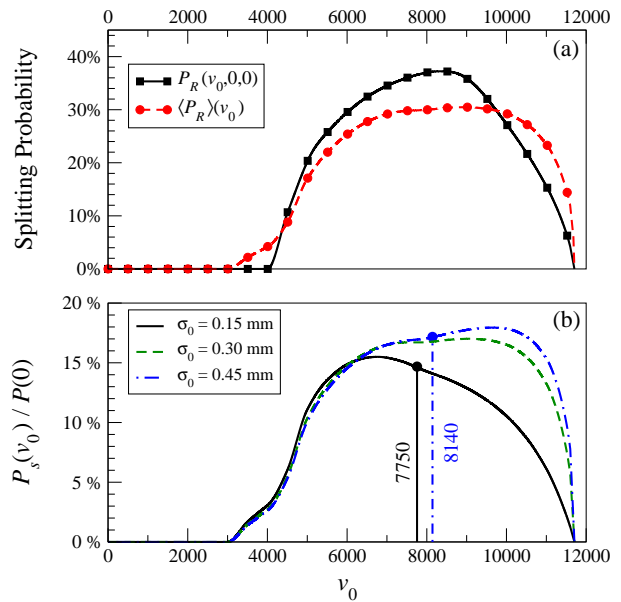}
\caption{(Color online) State-dependent splitting efficiency as a function of the initial level $v_0$. Upper graph (a)\,: the probability $P_R(v_0,0,0)$ of finding the atom in the oblique guide at the end of the propagation is shown as a black solid line for the initial condition \mbox{$z_0=\dot{z}_0=0$}. The same probability averaged over $z_0$ and $\dot{z}_0$, $\langle{P}_R\rangle(v_0)$, is shown as a red dashed line. Lower graph (b)\,: The averaged splitting probability $P_s(v_0)$ scaled by the same factor $P(0)$ as in Figure~\ref{fig:Prob_v} is shown as a function of $v_0$ for various sizes of the atomic cloud $\sigma_0$ and for \mbox{$T_0=14\,\mu$K}. The cloud sizes $\sigma_0=0.15$, 0.30 and 0.45~mm correspond to the solid black, dashed green and dash-dotted blue curves respectively. These probabilities have been averaged over the initial classical conditions chosen for $z_0$ and $\dot{z}_0$. The guide parameters are as in Figure~\ref{fig:Quantic}. The two vertical lines indicate the average value of $v_0$ in the oblique guide for $\sigma_0=0.15\,$mm (black solid line), $\left<v_0\right>=7750$, and $\sigma_0=0.45\,$mm (blue dash-dotted line), $\left<v_0\right>=8140$.}
\label{fig:Ps_v0_sigma0}
\end{figure}

The probability $P_R(v_0,z_0,\dot{z}_0)$ of finding the atom in the right wing potential well and thus in the oblique guide at the height of the detection probe ($z_p=-10\,$mm) is evaluated for each initial trap state $v_0$ by
\begin{equation}
\label{PRv0}
P_R(v_0,z_0,\dot{z}_0) = \int_{\rm Oblique \atop \rm Guide} \left|\varphi_{z_0,\dot{z}_0}(x,t_f)\right|^2\,dx\,.
\end{equation}
The wavepacket $\varphi_{z_0,\dot{z}_0}(x,t)$ is labelled here by the indexes $(z_0,\dot{z}_0)$ indicating the initial conditions of the simulation. The probability $P_R(v_0,0,0)$ is shown as a black solid line in the upper part of Figure~\ref{fig:Ps_v0_sigma0} as a function of $v_0$ for an atom initially at rest $(\dot{z}_0=0)$ at the height $z_0=0$. Clearly, the lowest energy states $(v_0 \leqslant 4000)$ are not deviated by the oblique guide, and they simply fall vertically. The explanation for this effect is simple: their energy is too small for them to be trapped in the present oblique guide of depth $U_1=10\,\mu$K (see for instance the lower inset of Figure~\ref{fig:Schematic} representing the guiding potential around \mbox{$z\simeq-8\,$mm}). The eigenstates of vibrational quantum number higher than \mbox{$v_0 \simeq 5800$} are the only states of total energy $\varepsilon_0 \geqslant -10 \mu$K. In an energy-based first approximation, all states with $v_0 \leqslant 5800$ should therefore remain unaffected by the beam splitter. In reality, all vibrational states experience a quickly varying potential in the vicinity of $z_c$. They are therefore subjected to non-adiabatic transitions to higher or lower excited states which may or may not be captured in the oblique guide. This non-adiabatic effect is especially important for the highest falling speeds, and thus for the initial conditions $z_0 > 0$ and \mbox{$\dot{z}_0 \neq 0$}. The upper part of Figure~\ref{fig:Ps_v0_sigma0} therefore shows the same probability but incoherently averaged over the initial distributions of $z_0$ and $\dot{z}_0$
\begin{equation}
\label{PRav0}
\langle{P}_R\rangle(v_0)  =  \int\!\!\!\!\int W_Q(z_0)\,W_{P}(\dot{z}_0)\,P_R(v_0,z_0,\dot{z}_0)\,\,dz_0\,d\dot{z}_0\,.
\end{equation}
In this figure, we see that the states of vibrational quantum numbers $3000 \leqslant v_0 \leqslant 5800$ already have a significant probability of splitting.  A comparison of the averaged probability distribution $\langle{P}_R\rangle(v_0)$ (red dashed curve in the upper part of Figure~\ref{fig:Ps_v0_sigma0}) with the probability $P_R(v_0,z_0,\dot{z}_0)$ obtained for a single initial condition ($z_0=\dot{z}_0=0$, black solid curve in the upper part of Figure~\ref{fig:Ps_v0_sigma0}) shows that even if the averaging procedure modifies significantly the probability distribution, a qualitatively correct description is already obtained by a single calculation with the atom initially at rest, with $z_0=\dot{z}_0=0$.

The initial position and momentum distributions of the atoms along $x$ is taken into account by balancing the averaged probability $\langle{P}_R\rangle(v_0)$ with the initial trapping probability $P(v_0)$. The splitting efficiency of the state $v_0$ is therefore written as
\begin{equation}
\label{Psv0}
P_s(v_0) = P(v_0)\,\times\,\langle{P}_R\rangle(v_0)\,.
\end{equation}
This state-dependent splitting probability $P_s(v_0)$ is shown in the lower part of Figure~\ref{fig:Ps_v0_sigma0} as a function of $v_0$ for various sizes of the atomic cloud $\sigma_0$ at fixed temperature \mbox{$T_0=14\,\mu$K}. An interesting tendency can be noticed in this figure~: larger atomic clouds, since they favor the initial trapping of higher vibrational levels in the vertical guide (see Figure~\ref{fig:Prob_v} for instance), have a higher total splitting efficiency, and present a distribution of levels clearly shifted to higher energies. As a consequence, the average value $\left< v_0 \right>$ of the trapped states is \mbox{$\left< v_0 \right> \simeq 7750$} for $\sigma_0=0.15\,$mm and \mbox{$\left< v_0 \right> \simeq 8140$} for $\sigma_0=0.45\,$mm.

The ``total" splitting efficiency $P_s$ at temperature $T_0$ is finally evaluated by averaging over the vibrational quantum numbers $v_0$ according to
\begin{equation}
\label{Ps}
P_s = \frac{1}{P_{\rm trap}} \sum_{v_0} P_s(v_0)\,,
\end{equation}
where $P_{\rm trap}$ is the total trapping probability [Eq.(\ref{Ptrap2})]. A unit splitting probability \mbox{($P_s = 1$)} would indicate that all trapped atoms are captured by the oblique guide. A perfect beam splitter, whose reflection and transmission coefficients equal 0.5, corresponds to \mbox{$P_s = 0.5$}.

\begin{figure}[t!]
\centering
\includegraphics[width=8cm,angle=0,clip]{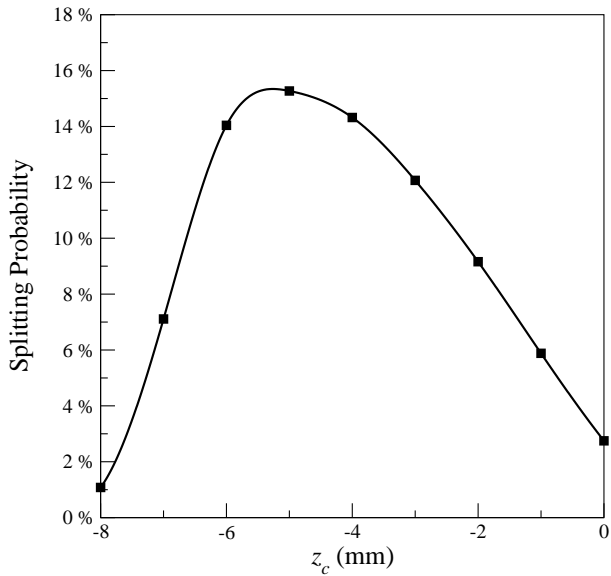}
\caption{Total splitting efficiency $P_s$ [Eq.(\ref{Ps})] as a function of the crossing height $z_c$ of the laser beams. This probability has been averaged over the initial conditions chosen for $z_0$ and $\dot{z}_0$, and summed over $v_0$. The size of the initial atomic cloud is $\sigma_0=0.30\,$mm. All other parameters are as in Figure~\ref{fig:Ps_v0_sigma0}.}
\label{fig:Ps_zc}
\end{figure}

The total splitting efficiency $P_s$ of the present beam splitter setup has been measured recently in Orsay for various heights of the crossing point $z_c$~\cite{Houde_PhD02}. The conclusion of this experimental study is that, with the parameters chosen in Figure~\ref{fig:Ps_zc}, a maximum splitting efficiency of about 10\% is observed around $z_c \simeq -6\,$mm. Some measurements made with a smaller waist $w_1$ and a higher potential depth $U_1$ also show that the variation of $P_s$ with $z_c$ is not symmetric with respect to its maximum value $z_c \simeq -6\,$mm. Our numerical study, which gives a maximum splitting probability of about 15\% for the height $z_c \simeq -5.2\,$mm (see Figure~\ref{fig:Ps_zc}) is therefore in qualitative agreement with this experimental measurement.

The variation of the splitting probability with $z_c$ can be explained as follows. If the average position of the atomic cloud is much lower than the crossing height when the oblique guide is switched on \mbox{($z_{ff}(t_0) \ll z_c$)}, the beam splitter is inefficient, as can be seen in Figure~\ref{fig:Ps_zc} for \mbox{$z_c \gg -4\,$mm}. On the opposite side of this graph, for \mbox{$z_c \ll -4\,$mm}, the oblique guide is on when the atoms reach $z_c$, but they reach this height with a kinetic energy which becomes comparable to -- or higher than -- the binding energy of the oblique guide $U_1$. This explains why the efficiency of the beam splitter falls to 0 when \mbox{$z_c \ll -4\,$mm}.

\begin{figure}[t!]
\centering
\includegraphics[width=8cm,angle=0,clip]{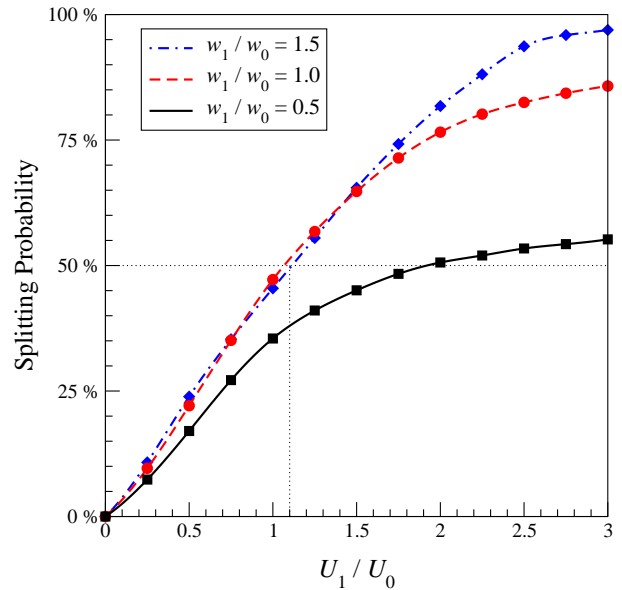}
\caption{(Color online) Total splitting efficiency $P_s$ [Eq.(\ref{Ps})] for the initial conditions $z_0=\dot{z}_0=0$ as a function of the ratio of the oblique to vertical potential depths $U_1/U_0$, and therefore as a function of the ratio of the beam intensities. For this calculation, $U_0$ is fixed (30~$\mu$K) and $U_1$ is varied. The crossing height between the two guides is $z_c=-4\,$mm. The waist of the vertical beam is $w_0=0.2\,$mm, and the splitting efficiencies calculated for an oblique waist of $w_1=0.1$, 0.2 and 0.3~mm are shown as black solid, red dashed and blue dash-dotted lines respectively. All other parameters are as in Figure~\ref{fig:Ps_zc}.}
\label{fig:Ps_U1_U0}
\end{figure}

We have also calculated the variation of the total splitting efficiency $P_s$ [Eq.(\ref{Ps})] of this beam splitter with one of the most crucial parameter~: the potential depth of the oblique guide $U_1$. This numerical simulation has been performed for various ratios of oblique to vertical beam waists $w_1/w_0$. The result is shown in Figure~\ref{fig:Ps_U1_U0} with fixed initial conditions $z_0=\dot{z}_0=0$. One can notice here that the splitting efficiency varies monotonically from 0 to its maximum value when $U_1$ varies from 0 to $3\,U_0$. Depending on the value of the waist of the oblique guide, a total deflection of the beam can be realized (see for instance the case $w_1 = 1.5\,w_0$ and $U_1 = 3\,U_0$). A completely symmetric splitting is also predicted when $U_1 \simeq 1.1\,U_0$ and $w_1 \geqslant w_0$. This last prediction is in agreement with the experiment~\cite{Houde_PhD02}.

The dash-dotted blue curve of Figure~\ref{fig:Ps_U1_U0} corresponds to a ratio of laser waists ($w_1/w_0=1.5$) very close to the experimental one~\cite{Houde00}. Our semi-classical model reproduces in this case the experimental splitting efficiency of 44\% (see Figure~4-f of reference~\cite{Houde00}) when $U_1/U_0=0.95$. A calculation performed for the same ratio of potential depths but with $z_c=-2\,$mm gives a splitting efficiency of 28.3\%, again very close to the experimental value of 29.2\% (see Figure~3-b of reference~\cite{Houde00}).

Finally, when the oblique guide is deep enough to induce a significant splitting of the atomic cloud, a higher splitting efficiency can always be obtained by increasing $w_1$. The results shown in this figure therefore indicate that a high degree of control exists in this type of experimental configuration since the splitting efficiency can be modified at will.

\begin{figure}[t!]
\centering
\includegraphics[width=8cm,angle=0,clip]{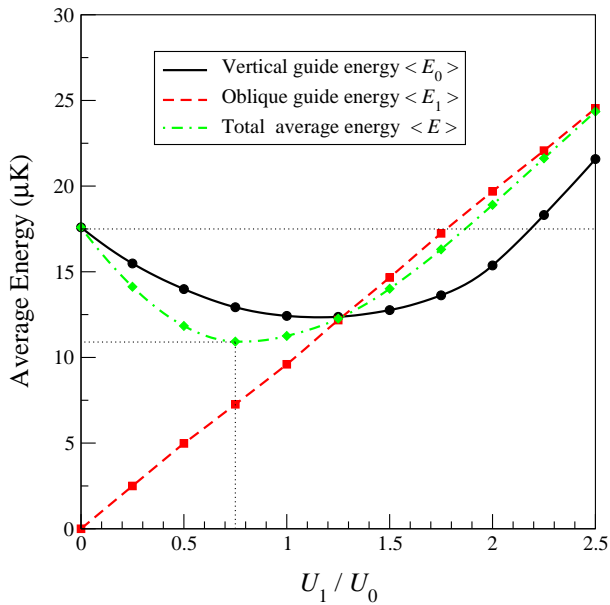}
\caption{(Color online) Average energies $\left\langle E_0 \right\rangle$ (black solid line) and $\left\langle E_1 \right\rangle$ (red dashed line) in the vertical and oblique guides as a function of $U_1/U_0$ for the initial conditions $z_0=\dot{z}_0=0$. For this calculation, $U_0$ is fixed (30~$\mu$K) and $U_1$ is varied. The crossing height between the two guides is $z_c=-4\,$mm. The waist of the vertical and oblique beams are $w_0=0.2\,$mm and $w_1=0.3\,$mm. All other parameters are as in Figure~\ref{fig:Ps_zc}. The total average energy $\left\langle E \right\rangle$ is also shown as a green dash-dotted line.}
\label{fig:E_U1_U0}
\end{figure}

Figure~\ref{fig:E_U1_U0} shows the average transverse energy (directions $x$ and $x'$) of the atoms in the vertical and in the oblique guide after the splitting~: $\left\langle E_0 \right\rangle$ and $\left\langle E_1 \right\rangle$. An evaluation of the final energy $E_0(v_0)$ in the vertical guide is first performed for each initial state $v_0$ using the expression
\begin{equation}
\label{eq:E0_v0}
E_0(v_0) = \int_{\rm Vertical \atop \rm Guide}
           \varphi_{v_0}^{*}(x,t_f)\,\hat{\cal H}_{\rm 1D}(x,t)\,\varphi_{v_0}(x,t_f)\,dx
\end{equation}
This energy is then averaged over all vibrational levels 
\begin{equation}
\label{eq:E0}
\left\langle E_0 \right\rangle =
\frac{\sum_{v_0} P(v_0)\,E_0(v_0)}{\sum_{v_0} P(v_0)\,P_0(v_0)}\,,
\end{equation}
where
\begin{equation}
\label{eq:P0_v0}
P_0(v_0) = \int_{\rm Vertical \atop \rm Guide}
           \varphi_{v_0}^{*}(x,t_f) \; \varphi_{v_0}(x,t_f)\,dx
\end{equation}
is the probability of experiencing a simple vertical fall when starting in the initial level $v_0$. In the oblique guide, a similar approach is used to calculate the average energy $\left\langle E_1 \right\rangle$, but the transverse direction is now $x'$. A rotation of the reference frame is therefore in order. For this calculation, the wavefunction $\varphi_{v_0}(x,t_f)$ and the Hamiltonian $\hat{\cal H}_{\rm 1D}(x,t)$ in Eq.(\ref{eq:E0_v0}) are thus replaced by
\begin{equation}
\label{eq:phibar}
\left\{
\begin{array}{lcl}
\widetilde{\varphi}_{v_0}(x',t_f) & \equiv & {\rm e}^{i\,\left[m\,\dot{z}(t)\,\tan\gamma\right]\,x} \; \varphi_{v_0}(x',t_f)\\
\widetilde{\cal H}_{\rm 1D}(x',t) & \equiv & \hat{\cal H}_{\rm 1D}(x',t) + m\,g\,\sin\gamma\,x'
\end{array}
\right.
\end{equation}

In parallel with the vertical and oblique average energies, Figure~\ref{fig:E_U1_U0} also shows the total average energy $\left\langle E \right\rangle$ of the trapped atoms after the splitting. This quantity is calculated from $\left\langle E_0 \right\rangle$, $\left\langle E_1 \right\rangle$, and the total splitting probability $P_s$ [Eq.(\ref{Ps})]
\begin{equation}
\label{eq:Emoy}
\left\langle E \right\rangle = \left(1-P_s\right) \left\langle E_0 \right\rangle
                               + P_s \left\langle E_1 \right\rangle\,.
\end{equation}

For $U_1=0$, no deviation of the cloud is observed, and we obtain $P_s=0$ and \mbox{$\left\langle E \right\rangle = \left\langle E_0 \right\rangle$}. This average transverse energy is in fact equal to the initial transverse energy of the trapped atoms (17.5\,$\mu$K). This behavior can be seen on the left part of Figure~\ref{fig:E_U1_U0}.

When $U_1$ increases by a small amount ($U_1 \leqslant U_0$) the highest vibrational levels of the vertical guide are deviated in the oblique potential (see Figure~\ref{fig:Ps_v0_sigma0} for instance), and the average transverse energy of the atoms remaining in the vertical guide therefore decreases. A striking counter-intuitive effect is that the atoms which are deviated also have a translational energy which is smaller than the initial average energy of the trapped atoms. This happens because these high vibrational levels are now trapped in a weakly binding potential of depth $U_1 \leqslant U_0$. As a consequence, the total average translational energy $\left\langle E \right\rangle$ of the atoms in their transverse direction decreases after the splitting of the cloud. With the parameters used in Figure~\ref{fig:E_U1_U0}, a minimum energy of 10.9\,$\mu$K is obtained for $U_1 \simeq 0.75\,U_0$, to be compared with the initial average energy of about 17.5\,$\mu$K. A significant cooling effect is therefore obtained in the transverse direction, at the cost of a significant heating in the vertical direction.

Finally, on the right hand side of this figure, with very deep oblique potentials \mbox{($U_1 \geqslant 2\,U_0$)}, the atom temperature in the transverse direction exceeds the initial average energy of 17.5\,$\mu$K. In this case, a heating process takes place due to the fact that the atoms are now trapped in a much deeper potential.

\section{Conclusion}
\label{sec:Conclusion}
We have proposed a theoretical model for the study of a thermal ensemble of cold atoms in a beam splitter device. This model has a wide range of possible applications. For instance it could be used to describe the dynamics of cold atoms trapped and manipulated with the magnetic fields created by atom chips. We have used our time-dependent semi-classical model to describe the atomic dynamics in the presence of two crossing dipole guides. We have taken into account the gravity, as well as the thermal population of the initial atomic cloud in order to compute the splitting efficiency of the beam splitter.

Our results are in good agreement with experimental measurements, and we have presented the influence of the main parameters on the atomic dynamics in this guiding and splitting configuration. We have shown that some eigenstates of the system split coherently in the two branches of the guide, and that different average temperatures can be obtained in the different arms of the beam splitter. An efficient cooling of the atoms is also predicted in the transverse direction.

All these results indicate that a high degree of control can be achieved in this type of cold atom beam splitters, using simple Gaussian laser beams. In the future our investigations will concentrate on the theoretical description of atom optics devices (guides, mirrors,~\ldots) for the manipulation of both thermal and coherent sources of atoms. Combined with the very impressive capabilities of spatial light modulators~\cite{McGloin03}, these techniques should effectively allow for the implementation of new exciting experimental schemes in the domain of atom optics and matter-wave interferometry.

\begin{acknowledgments}
We thank Herv\'e Le Rouzo for stimulating and helpful discussions. The IDRIS-CNRS supercomputer center supported this study by providing computational time under project number 08/051848. This work has been done with the financial support of the LRC of the CEA, under contract number DSM 05--33.  Laboratoire de Photophysique Mol\'eculaire and Laboratoire Aim\'e Cotton are associated to Universit\'e Paris-Sud 11.
\end{acknowledgments}

\end{document}